\documentstyle[11pt]{article}
\input{epsf}

\def\hybrid{\topmargin -20pt    \oddsidemargin 0pt
        \headheight 0pt \headsep 0pt
        \textwidth 6.5in        
        \textheight 9in         
        \marginparwidth .875in
        \parskip 5pt plus 1pt   \jot = 1.5ex}

\newskip\humongous \humongous=0pt plus 1000pt minus 1000pt
\def\caja{\mathsurround=0pt}
\def\eqalign#1{\,\vcenter{\openup1\jot \caja
        \ialign{\strut \hfil$\displaystyle{##}$&$
        \displaystyle{{}##}$\hfil\crcr#1\crcr}}\,}
\newif\ifdtup

\relax
\hybrid

\def\be{\begin{equation}}
\def\ee{\end{equation}}

\newcommand{\beq}{\begin{equation}}
\newcommand{\eeq}[1]{\label{#1}\end{equation}}
\newcommand{\ber}{\begin{eqnarray}}
\newcommand{\eer}[1]{\label{#1}\end{eqnarray}}

\def\CD{{\cal D}}
\def\quarter{{1 \over 4}}
\def\pd{\partial_1^{ }} \def\pdd{\partial_1^2}
 
\def\po{\partial_0^{ }} \def\half{{1\over2}}
\def\IL{{\rm \relax{I\kern-0.18em L}}}
\def\IH{{\rm \relax{I\kern-0.18em H}}}
\newcommand{\newsec}[1]{\section{#1}}
\newcommand{\subsec}[1]{\subsection{#1}}
\def\eqn#1#2{\beq #2 \eeq{#1}}
\def\exclude#1{}

\begin{document}
\overfullrule=0pt

\def\baselinestretch{1.2}
\baselineskip 16 pt

\begin{titlepage}

\begin{flushright}
RU-98-09\\
\end{flushright}
\vspace{10 mm}

\begin{center}
{\huge Classical Decay of the Non-SUSY-Preserving 
Configuration of Two D-Branes.}

\vspace{5mm}

\end{center}

\vspace{5 mm}

\begin{center}
{\large Anatoly V. Morosov }

\vspace{3mm}

Department of Physics and Astronomy\\
Rutgers University\\
Piscataway, New Jersey 08854-0849
\end{center}

\vspace{1cm}

\begin{center}
{\large Abstract}
\end{center}

\noindent
We have studied a problem of the tachyon mediated D-brane - D-brane 
annihilation from the underlying world-volume gauge field theory
point of view.
The initial state was chosen in the form of two
D-branes crossing at a non-zero angle, which is 
non-supersymmetric configuration generically. 
This state was not a groundstate of the theory and the problem 
at hand was a model, where we had rather precise control over 
the behavior of the theory.
Some applications of this model to the D-brane physics and conclusions about
stability of several configurations were made.
By taking a T-dual picture of this process on a $T^2$ torus, we derived 
known conclusion about the
instability of a system consisting of D-0 and D-2 branes, and found
its decay modes.

\vspace{4cm}
\begin{flushleft}
March 1998
\end{flushleft}
\end{titlepage}
\noindent
 \setcounter{equation}{0}

\newsec{Introduction.}

The Dirichlet $p$-branes in the type II superstring theories are defined
as objects, on which the fundamental strings can end \cite{D orig}. 
The low energy dynamics of D-branes comes from massless modes 
in the open-string sector 
of a worldsheet theory. In the case of a flat supergravity background, it
is given by a gauge field theory
dimensionally reduced from 10 dimensions down to $p+1$ \cite{witten,polch} 
\footnote{for reviews see \cite{polch2, bachas, taylor}}.
For slowly varying fields this theory can be approximated
by a world-volume Dirac-Born-Infeld theory with a gauge group U($n$),
and for small perturbations around flat D-brane surface it can be 
approximated further by a U($n$) 
super-Yang-Mills theory. We will use the latter approximation, 
although it is discovered lately that there are
important phenomena missing in the Y-M approximation. 
For example, the 
process in which a fundamental string emanates from a D-brane
(cf. \cite{CM, LPT}).

The translation between the languages of  the (super-)Yang-Mills theory
on a world-volume and the perturbative string theory with D-branes 
included can be done in the following manner.
If in the language of a 10-dimensional U($n$) Yang-Mills in some 
particular gauge, we have a solution for
a vector field ${\hat A}_\mu$
\be
{\hat A}_{\mu}=\half\pmatrix
{A_\mu^{(1)}(x_0...x_p)&0&...&0\cr
0&A_\mu^{(2)}(x_0...x_p)&...&0\cr
...\cr
0&0&...&A_\mu^{(n)}(x_0...x_p)\cr},
\ee
then we say, that  ${1\over2\pi\alpha'}A_i^{(k)}(x_0...x_p)$ 
define transverse positions 
of $n$  D$p$-branes for $i=p+1,..,9$, and $A_a^{(k)}(x_0...x_p)$ for
$a=0..p$ is just a world-volume gauge field, living in a D-brane $(k)$.
All the off-diagonal excitations around this configuration, like 
$A_\mu^{i,j}$, correspond to the fundamental open strings stretched from 
the brane $(i)$ to the brane $(j)$.

As an example of using this dictionary, one can consider a D-string
twice wound around the circle $S^1$. In the language of the gauge fields
 it corresponds to a twisted boundary condition on a circle,
which permutes diagonal elements of a U(2) matrix.

There are different configurations of D-branes preserving some number of 
supersymmetries. One D-brane, for example, breaks down one half of all the 
supersymmetries. Other known configurations can 
preserve other part of supersymmetries, like 1/4 or 3/8 
(\cite{angles, townsend}).
Any D-brane configuration preserving at least some of supersymmetries is a
true groundstate of the theory.

In this paper we describe in some details the behavior of a 
system consisting of two
D-strings crossing at a non-zero angle (parameter $\theta$ is a tangent 
of the crossing angle) 
and separated in one more spatial direction by the distance $a$.
This configuration breaks down all the supersymmetries so it is not 
a groundstate and can decay into something else.
We found that this configuration indeed was unstable 
when $\theta -2\pi\alpha' a^2>0$.
The decay mode of this system can be effectively described in the string theory
language as a process which intercommutes the ends of D-strings. The bent
string emerging as a result moves away from the crossing point.
We found that the characteristic time of this decay is of order 
$\tau\approx l_s(\theta - a^2/l_s^2)^{-\half}$.
By taking this theory on a $T^2$ and by T-dualizing it,
we find similar conditions for unstability of a D2-D0 system, when the angle
between two D-strings translates into the complex structure of a $\tilde{T}^2$.

\newsec{D-strings crossing.}

The model we have studied was the configuration of two D-strings, 
which are separated along $x^2$ coordinate by distance $a$, 
and are rotated one relatively to another
in the $(x^1,x^3)$ plane by a non-zero angle.
Everywhere in the text we use the parameter, $\theta$ which is a tangent 
of a crossing angle between these two D-strings.
The static force between two crossing D-strings was
calculated using string technics in one loop order in \cite{alec}.

We would like to emphasize, that since the configuration considered
is not BPS, the spectrum will have quantum corrections.

{\bf Conventions:} In this paper we use dimensionless units, by taking 
${1\over2\pi\alpha'}=l_s^2=1$ and the 
signature is taken (-,+,...+). Greek indices $\mu,\nu...$ are 10-dim indices,
$a,b...$ are world-volume indices or SU(2) group indices, depending 
on a context 
and $i,j...$ are indices in the directions transverse to a D-brane.

Using the property that U(1) subgroup of U(2) corresponds
to the center of mass motion and decouples, we will use SU(2) as the 
gauge group of the theory.

\subsec{First approach. Off-diagonal excitations around 
two-strings-crossing configuration.}

In this work we consider unstable
classical solutions containing bosonic fields only.
Our starting point was a 10-dim Yang-Mills SU(2) vector field $A_{\mu}$, with 
the bosonic part of the Lagrangian density given
by 
\be
{\cal L}=-\half {\rm Tr} F_{\mu\nu}F^{\mu\nu}.
\label{lagr}
\ee

Let's consider excitations around field configuration, corresponding
to two D-strings crossing at some non-zero angle:
\be
{\hat A}_{\mu}=\half\pmatrix{a_\mu&0\cr0&-a_\mu\cr},
\ee
where $a_{\mu}=(0,0,a,\theta x,0..0)$, where plain $x$ will 
denote a spatial coordinate $x_1$ along the D-string. 
We have considered the case with D-strings only, so the value 
of $p=1$ is taken.

In this chapter we will focus on modes corresponding to a string, 
stretched between branes ``1'' and ``2'' 
only\footnote{It means we will not consider the motion of a brane 
itself for awhile.},
 so we take perturbations in the form
\be
\delta {\hat A}_\mu = \half\pmatrix{0&A_\mu\cr A^*_\mu&0\cr},
\ee
It turns out that the tachyon excitation lives in this sector.

Performing dimension reduction, we get the 2-dimensional theory with
``mixing'' for fields $A_0,\ A_1,\ \Phi_2$ and $\Phi_3,$ and $x$-dependent
``mass'' term of form $m^2=a^2+\theta^2x^2$ for every other field $\Phi_k$ 
in this problem.  The eigenstates and eigenvalues of the Lagrangian 
$x-$differential operator we got, correspond to the masses we would 
have had, if we 
had no explicit $x-$dependence in this operator.

\subsec{Lagrangian and Hamiltonian formulation of the world-volume theory.}

\exclude{
\be
\eqalign{& {\cal L}=
 \half(\po A_1^{ }-\pd A_0^{ })(\po A_1^*
-\pd A_0^*)+
\half\po\Phi_{2..9}^{ }\po\Phi_{2..9}^* 
-\half\pd\Phi_{2..9}^{ }\pd\Phi_{2..9}^*\cr
& +{i\theta\over2}(-A_1^{ }\Phi_3^*+\Phi_3^{ }A_1^*)
+\half(a^2+\theta^2x^2)(A_0^{ }A_0^*-A_1^{ }A_1^*)\cr
& -\half\theta^2x^2\Phi_2^{ }\Phi_2^*-\half a^2\Phi_3^{ }\Phi_3^* 
 + {a\theta x\over2}(\Phi_2^{ }\Phi_3^*+\Phi_2^*\Phi_3^{ }) 
 -\half(a^2+\theta^2x^2)\Phi_{4..9}^{ }\Phi_{4..9}^* \cr
& -{i\over2} (a A_0^{ } (\po\Phi_2^*)-a A_1^{ }(\po\Phi_2^*) 
+\theta xA_0^{ }(\po\Phi_3^*)-\theta xA_1^{ }(\pd\Phi_3^*)) \cr
& -{i\over2} (-a(\po\Phi_2^{ })A_0^* + a(\pd\Phi_2^{ })A_1^*
 -\theta x(\po\Phi_3^{ })A_0^* 
+ \theta x(\pd\Phi_3^{ })A_1^*)\cr
& +{1\over16} (A_{[\mu}^{ }A_{\nu]}^*)^2}
\ee
}
It turns out that the search for a static spectrum is easier to perform in the
Hamiltonian formulation of that theory. We will not fix any gauge freedom
present in the Hamiltonian, since we are considering classical excitations
and any allowed gauge transformations can be undone later, if one needs it, 
in terms of the classical solutions.

\def\FieldsL{\pmatrix{A_1^*&\Phi_2^*&\Phi_3^*\cr}}
\def\FieldsR{\pmatrix{A_1\cr\Phi_2\cr\Phi_3\cr}}

\def\FirstM{
\pmatrix{a^2+x^2\theta^2&ia\pd&ix\theta\pd-i\theta
\cr ia\pd & -\pdd+ x^2\theta^2 & -ax\theta 
\cr ix\theta\pd+2i\theta  & -ax\theta  & -\pdd+a^2 \cr  }}

The canonical momenta conjugated to the fields 
$A_{0,1},\ \Phi_{2..9}$ are $\pi_{0..9}$ and the Hamiltonian for this system
looks like

\be
\eqalign{
& {H}=\int dx \big(
      2\pi_{1..9}^*\pi_{1..9}^{ }
      +(A_0^{ }(-\partial_1\pi_1+ia\pi_2+i\theta x\pi_3)+c.c.)\cr
& +\half\FieldsL\FirstM\FieldsR
-{1\over16}(A^{ }_{[\mu}A^*_{\nu]})^2\big)\cr}
\ee

Let 
\be
\nonumber
{\IH=\FirstM}
\ee

From the {\it bona fide} quadratic ($x^2$) behavior of the ``potential'' 
(as well as from the
point of view of a stretched string) we would expect the 
normal modes of a differential operator $\IH$ to be the oscillatory
ones. This ansatz indeed gives us the lowest normalizable static 
eigenstate as:
\be
{T\propto\pmatrix{i\cr0\cr 1}e^{-{\theta x^2 \over 2}},}
\ee
with $\IH T=(a^2-\theta)T$, and it becomes tachyonic if $a^2<\theta$.

The second lowest eigenvalue of this operator is $(a^2+\theta)$, it can not 
become tachyonic, so it can not lead to any instability.

\subsec{$T^4$ member of Lagrangian. ``Stable'' tachyonic vev.}

Since the eigenstates $(V_k)$ of $\IH$ form a complete basis in 
the Hilbert space $H^3$, we can write arbitrary $A_1$ and $\Phi_{2..9}$ as 
\ber
& A_1(t,x)=\Sigma_k Z_{1,k}(t) V_k(x)=
- i Z_{0}(t)e^{-{\theta x^2 \over 2}}+
\Sigma_{k\neq 0} Z_{1k}(t) V_k(x)\\
& \Phi_3(t,x)=\Sigma_k Z_{3,k}(t) V_k(x)=
Z_{0}(t)e^{-{\theta x^2 \over 2}}+
\Sigma_{k\neq 0} Z_{3k}(t) V_k(x)\\
& \Phi_i(t,x)=\Sigma_{k\neq 0} Z_{i,k}(t) V_k(x),\ \ i=2,4..9
\eer{rever}

Of all the original fields, only $A_1$ and $\Phi_3$ contain a tachyon.
Supposing that no other fields are excited, and we are interested in 
the terms in the Hamiltonian containing $Z_0$ only, then we can 
substitute (\ref{rever}) into $H$ and integrate over $x$.
We get the following ``potential'' for (static) $Z_0$:
\beq
{H=
\sqrt{\pi\over\theta}\left((a^2-\theta)Z_0^*Z_0^{ }
+{1\over2\sqrt{2}}(Z_0^*Z_0^{ })^2 \right)+...}
\eeq{forZ}

So when $a^2-\theta<0$ we have the symmetry spontaneously broken, and 
even if we have $Z_0(t=0)=0$, it evolves with time to a non-zero value:
\eqn{vevZ}{\langle Z_0^*Z_0^{ }\rangle=-{(a^2-\theta)}\sqrt{2}}
Assuming $\langle Z_0\rangle$ real we get the following  
stable static 
vevs\footnote{Under the term {\it vev} we assume just a classical solution
for the equations of motion.} for the initial fields
\eqn{vevA}{\langle A_1\rangle =-i\sqrt{-(a^2-\theta)\sqrt{2}}
e^{-{\theta x^2\over2}},}
\eqn{vevF}{\langle\Phi_3\rangle=\sqrt{-(a^2-\theta)\sqrt{2}}
e^{-{\theta x^2\over2}},}
and no other fields evolve to a non-zero static value due to this process.

\subsec{Decay of the initial configuration.}

The initial configuration corresponds to two D-strings crossed. Since 
at the initial moment of time the gauge field is zero, the Wilson's
line is trivial, so the D-strings in the original picture are
distinguishable entities. The
U(2) gauge field $A_1$ becomes non-zero as the time goes on, 
so that the 
\be
\nonumber
\langle{\hat A}_1\rangle=\half\pmatrix{0&\langle A_1(x)\rangle\cr
\langle A_1^*(x)\rangle&0\cr},
\ee
gives a non-trivial Wilson's line for the original (fixed in the 
given framework) diagonal $\Phi_3$ configuration
\eqn{Wil}{ U_p(\infty,-\infty)=Pexp(i\int{\rm d}x\langle\hat A_1(x)\rangle) 
=\pmatrix{\cos(c)& -\sin(c)\cr \sin(c)&\cos(c)\cr},}
where $c=\sqrt{\pi\over\theta\sqrt{2}}\sqrt{-(a^2-\theta)}$

We immediately see, that such a tachyon mediated evolution
mixes the ends of two original D-strings. They do not belong
to different distinguishable D-strings anymore.
A problem with that perturbative solution we found is that it does not exchange
ends completely ($c\neq\pi/2$), but just mixes them. This is
just an artifact of taking a perturbative approach and 
including off-diagonal excitations only. The decay of
the resulting system, consisting of two bent D-strings cannot be found 
in the framework of this chapter, since we did not consider any change
for diagonal part of $\hat \Phi$ matrix, which describes the positions
of two D-branes. The bent D-strings will obviously tend to straighten
themselves, so they effectively will move away from the crossing point. 
We will show
in the next chapter, that the running away solution is indeed a stable 
classical solution for a full non-perturbative interacting theory.

\newsec{Unstable and stable configurations.}

Since we have established a condition when the tachyonic mode arises: 
$a^2-\theta<0$,
(or in dimensionful quantities $a^2-l_s^2\theta<0$) we can restrict
ourselves to the simpler case of $a=0$ and we will not lose anything 
interesting.
The physics of the process with a tachyonic mode and $a\neq0$ 
is the same as the one considered below. Something different can happen
when we take a boundary value, $a^2-\theta=0$, and we might have
a classical perturbative flat direction, free for the system to move 
along. It may be unstable
when properly considered, with all the higher order terms included.

\subsec{Full interacting SU(2) theory. Unstable solution.}

\eqn{lagra}{{\cal L}=-{1\over 2}{\rm Tr}F_{\mu\nu}F^{\mu\nu}.}

After dimensional reduction to $1+1$ dimensions the Lagrangian 
becomes\footnote{Upper $a$ index is a SU(2) group index, running 1..3}:

\eqn{lagrb}{{\cal L}_R=\half(F_{01}^a)^2+\half(\CD_0\Phi_i^a)^2-
\half(\CD_1\Phi_i^a)^2-
\quarter(\epsilon^{abc}\Phi_i^b\Phi_j^c)^2}

Let's consider such classical field configurations only, which have
non-zero fields $A_1^2,\ \Phi_3^1,$ and $\Phi_3^3$, and all the
other fields are vanishing. As we have established in the previous 
chapter, those chosen $A_1^2,\ \Phi_3^1,$ are sufficient to 
describe a tachyon,
and the $\Phi_3^3$ will describe the string's position.

(N.B. that the field $A_0$ provides us with one equation of 
motion, the \ref{EOMs}(1) below, even vanishing by itself.)

To clarify notation we will use from now on the following symbols:
$$A\equiv A_1^2,\  \Phi_1\equiv\Phi_3^1,\  \Phi_3\equiv\Phi_3^3$$

The corresponding equations of motion are 
\eqn{EOMs}{\eqalign{
& (1)\ \ \ -\dot{A}'+\Phi_3\dot{\Phi}_1-\dot{\Phi}_3\Phi_1=0\cr
& (2)\ \ \ -\ddot{A}+\Phi_3^{ }\Phi_1'-\Phi_3'\Phi_1^{ }
      -A(\Phi_1^2+\Phi_3^2)=0\cr
& (3)\ \ \ -\ddot{\Phi}_1^{ }+\Phi_1''-A'\Phi_3^{ }
      -2A\Phi_3'-A^2\Phi_1^{ }=0\cr
& (4)\ \ \ -\ddot{\Phi}_3^{ }+\Phi_3''+A'\Phi_1^{ }+2A\Phi_1'^{ }
      -A^2\Phi_3^{ }=0\cr}}

The static solution, corresponding to the initial configuration
of two crossing 1-branes is given in our original gauge by: 
$$A=0,\ \Phi_1=0,\ \Phi_3=\theta x.$$ Let this gauge, where
the initial configuration looks this way be called {\bf Gauge-A}.

As we know from the previous chapter, this solution is unstable
relative to the perturbation of the form 
$\delta A=-\epsilon$, $\delta\Phi_1=\epsilon$. Now, we will show this 
explicitly using different methods.

\subsec{Gauge-A and Gauge-B.}
We will call {\bf Gauge-B} any gauge, where $\Phi_3$
defines bent strings. 
Any transformation between {\bf A}$\rightarrow${\bf B} is given by
a transformation non-trivial at one of the spatial infinities, while any
{\bf B}$\rightarrow${\bf B$'$} or {\bf A}$\rightarrow${\bf A$'$} 
is given by a gauge transformation
trivial at the spatial infinities.

For our initial configuration, it's obvious, that any {\bf Gauge-A}
has the trivial Wilson's line $U_p$, while any of {\bf Gauge-B} has purely 
off-diagonal $U_p$.

\begin{figure}
\begin{center}
\leavevmode
\epsfbox{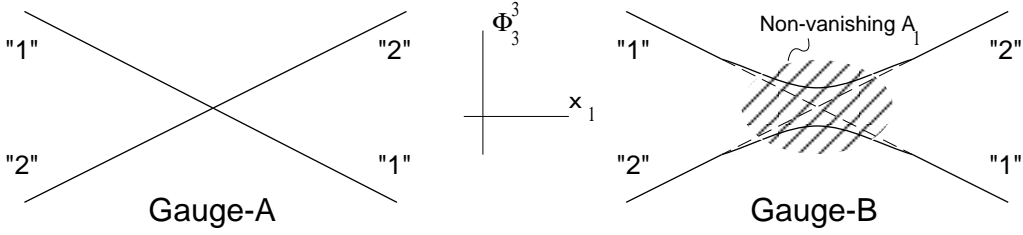}
\end{center}
\caption{
Initial configuration of D-strings in the Gauges A and B.
\label{diagram}}
\end{figure}

Local and time-independent SU(2) gauge transformation of form
$$g(x)=\pmatrix{f_1(x)& f_2(x)\cr-f_2(x)&f_1(x)\cr}$$ where
$$\eqalign{& (f_1(x))^2=\half(1+{\rm th}(\alpha x))\cr
& (f_2(x))^2=\half(1-{\rm th}(\alpha x))\cr},$$
with $\alpha$ -- is an arbitrary parameter, is an example of a gauge 
transformation, which takes us to the 
{\bf Gauge-B}. In this particular gauge, the initial static configuration 
of fields plus perturbation of our kind (in the Gauge-A: $A=-\epsilon$, $\Phi_1=\epsilon$, $\Phi_3=\theta x$)
looks like:
$$\eqalign{& A={\alpha\over {\rm ch}\alpha x}-\epsilon\cr
& \Phi_1=-{\theta x\over {\rm ch}\alpha x}+\epsilon\ {\rm th}\alpha x\cr
& \Phi_3=\theta x\ {\rm th}\alpha x+{\epsilon\over {\rm ch}\alpha x}\cr}$$

The initial
configuration in the {\bf Gauge-B} has the Wilson's line 
\eqn{Wila}{ U_p(\infty,-\infty)=Pexp(i\int_{-\infty}^{+\infty}dx\hat{A}(x)) 
=\pmatrix{0&1\cr-1&0\cr},} completely off-diagonal, as expected,
so these `bent'
strings are really cross-connected (as it was obvious in the {\bf Gauge-A}).

If we take $\epsilon$ of form $\epsilon(t,x)=e^{i\omega t}\epsilon(x)$,
we'll find that in the approximation linear in $\epsilon$, the unstable
solution with $\omega^2=-\theta<0$ exists and corresponds to 
$\epsilon(x)\propto e^{-\theta x^2/2}$. So this is indeed the direction 
of decay for our original configuration.
By taking $\epsilon(t=0,x)=0$ and slightly ``kicking'' this initial field
configuration, we will move along this downward direction until a new
(non-static) equilibrium is found.

Evolution of the Wilson's line in the {\bf Gauge-B} looks like
\eqn{Wilb}{ U_p(\infty,-\infty)=Pexp(i\int dx (\hat{A}(x)-\hat\epsilon(x))) 
\approx\pmatrix{\epsilon(t)&1-\epsilon(t)^2/2
\cr-(1-\epsilon(t)^2/2)&\epsilon(t)\cr},}
so the classical motion downward in this direction is going to mix the 
ends of D-strings.

\subsec{Elementary estimation of a decay rate for the scattering process.}

We have considered the following initial configuration (in
the {\bf Gauge-A}): 
\eqn{decay rate}{\Phi_3^3=\theta x,\ \ \Phi_2^3={\rm const}-vt,}
similar to the configurations discussed in the section (2), for the
D-strings crossing at some distance $a$, except that now the D-strings are
moving. This configuration satisfies equations of motion, so it is
an (unstable) solution.  To estimate this
instability for different values of $v$ - relative velocities of
D-strings, we notice, that the characteristic time of D-string decay,
as prompted by $\exp(\sqrt{\theta}t)$ behavior of $\epsilon(t)$ is
$\tau_d=\theta^{-\half}$ while the characteristic time of interaction
is $\tau_i=\theta/v$. So if $v\le\theta^{3\over2}$ these decay processes
are important.

{\bf NB:} The estimation can break down, since the decay rate 
$\tau_D=(\theta-a^2)^{-\half}$ was calculated in the assumption of 
static brane configuration.

\subsec{Stable final configurations.}

On the other hand, the final configuration (as $t\rightarrow\infty$), 
is supposed to have  $A=0$ in one of the {\bf Gauges-B}, and the equations 
of motion for the fields $A,\ \Phi_1,\ \Phi_3$ imply the wave
equations for fields $\Phi_1$ and $\Phi_3$
$$ -\ddot{\Phi}_{1,3}^{ }+\Phi_{1,3}''=0{\rm ,\ with\ relation}$$
$$\Phi_1={\rm const\ }\Phi_3.$$
So we can eliminate $\Phi_1$ from the final state completely 
by a {\it global} gauge transformation.

\begin{figure}
\begin{center}
\leavevmode
\epsfbox{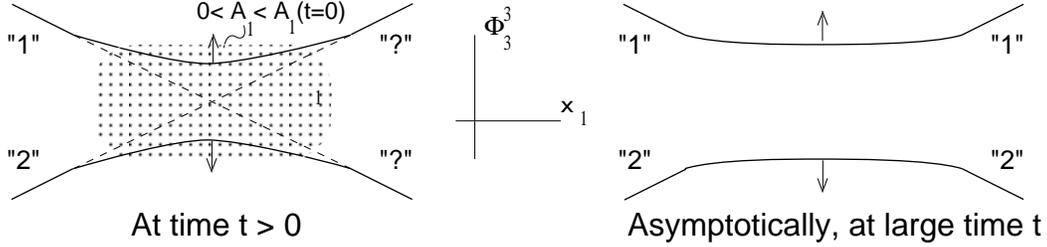}
\end{center}
\caption{Evolution of the initial configuration in the Gauge-B.
\label{diagram2}}
\end{figure}

To show that this configuration is stable, we need to notice only,
that for such an asymptotic wavelike behavior of $\Phi_3$, it becomes 
arbitrarily big
as time $t\rightarrow\infty$: i.e. the D-strings become widely separated,
and the separation increases with the time. 
$${\rm min}\|\Phi_3(t)\|\geq \theta t$$

The linearized equation of motion \ref{EOMs}(2) for a small 
$\delta A\propto e^{i\omega t}$ perturbation around such a solution 
looks like:
\eqn{waveA}{
\omega^2\delta A=-\Phi_3^{ }\delta\Phi_1'+\Phi_3'\delta\Phi_1^{ }+(\Phi_3^{ })^2\delta A,}
and the eventually huge positive part $(\Phi_3^{ })^2$ kills everything else. 
So there are no
``downward'' directions which can change $A$, but there is a
flat direction which leaves $A$ intact and moves $\Phi_1$ in the $\Phi_3$
direction. As was mentioned above, this can be undone by some global 
gauge transformation, which does not change anything physically.

\subsec{Open questions.}
Since the behavior of $\Phi_3(t)$ was not found explicitly
for the intermediate times, we cannot say for sure which of the
final states the system chooses, i.e. what is the final profile of a wave-like 
solution.

{\bf NB:} The Born-Infeld action was approximated by a Yang-Mills action
to consider the stability of a static solution, and it was a good
approximation -- i.e. for a small deviation from a flat D-brane surface. 
The Born-Infeld action is more appropriate for the purpose of describing a
running away and bent D-string. As we have seen above, interaction
between these runaway D-strings decreases as the separation increases,
and at some point we can neglect the interaction with the other D-string 
completely,
thus the Born-Infeld action becomes Nambu-Goto action, describing 
a lone runaway string.

\exclude{
Using the advantage of having just a 1+1-dim worldsheet, the Nambu-Goto
equations can be substituted with Polyakov's equations, and the resulting 
motion can be described by a wave equation with additional restrictions:
}

For a 1+1-dimensional worldsheet in the classical case the question 
will be very simple, it is just a choice of a proper parametrization 
for a coordinate
$\sigma$ along the string. We cannot choose $X_1=\sigma$ as before, but
taking a wavelike solution (in the light cone coordinates 
$x_+=\half(\sigma+\tau)$ and $x_-=\half(\sigma-\tau)$) in a form of 
\eqn{NambuGoto}{\eqalign{
& X_0=x_+-x_- \cr
& X_1=f_1(x_+)+f_1(x_-) \cr
& X_3=f_3(x_+)+f_3(x_-), \cr}}
with an extra-condition \eqn{extra}{f_1'^2+f_3'^2=1} we find that it satisfies 
the Nambu-Goto equations of motion, and (\ref{extra}) provides us inexplicit
expression for $\sigma(X_1(\tau=0))$ giving in principle 
the shape $X_3(X_1,\tau)$ as expected.

\begin{figure}
\begin{center}
\leavevmode
\epsfbox{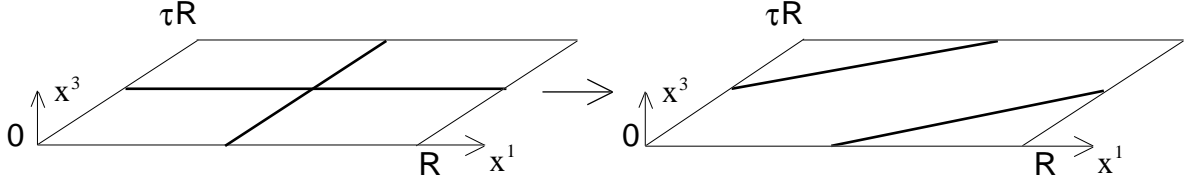}
\end{center}
\caption{Evolution of the initial configuration on the torus $T^2$.
\label{torus1}}
\end{figure}

\newsec{Crossing strings on a torus and a dual torus.}
\subsec{The toroidal compactification.}
We put the D-strings crossing on a torus $T^2$, which compactifies
dimensions $x^1$ and $x^3$ (cf. fig. \ref{torus1}).
Since we have considered the U(2) group only, it describes our initial 
configuration for two D-strings with a winding numbers {\bf 1+1} along 
the foundations
of a torus, so they are parallel to the sides of a torus.
\footnote{If we have one D-string wound more than once, say $n$ times, 
then the corresponding Yang-Mills theory, which is to describe this D-string, 
must have a gauge group U(N), with N not less than $n$. Otherwise we cannot 
embed any non-trivial
Wilson's line distinguishing between different turns. Alternatively, if we 
think about the theory in a small region far from the torus' borders, 
$n$-wound theory is a gauge U($n$) theory. So the case of U(2) describes
two single-wound D-strings (or one double wound), and this is possible 
for non-parallel D-strings only in the case when sides of a 
torus are parallel to our D-strings.}
Thus the method considered describes only tori with zero flux and such a
modular parameter $\tau$ so ${\rm Im\ }\tau/{\rm Re\ }\tau=\theta$.

The decay approximation is valid when the dimensions of the torus are much 
larger 
than the tachyon field size, $l_s\theta^{-1/2}$. The final configuration in 
the case of decay corresponds to one D-string with a winding number 
$1_1+1_3=2$ around this $T^2$.

\subsec{T-dual picture.}
If we consider a picture in a torus $\tilde{T}^2$, T-dual to original 
one along one direction (say $x^1$) only,
then we have the initial configuration of a D-2 brane (from a D-string 
along the $x^3$ on a $T^2$) and a D-0 brane (from a D-string wound
around the $x^1$).
From the previous discussion we are aware of the instability of this
configuration, we can conclude that in the final state D-0 brane 'dissolves' 
in a D-2 brane wrapped around  $\tilde{T}^2$ and, in principle, we can 
describe this process of dissolution. The final state can be best described
as once wrapped D-2 brane on a  $\tilde{T}^2$ with one unit of 
uniformely distributed flux (cf. \cite{taylor} ch.4.4).

Thus the combination of D-0 and D-2 branes compactified on tori are 
unstable\footnote{this fact can be easily understood in terms of 
energy minimization 
(cf. \cite{taylor} and ref. thereof)}. The
measure of the instability for this picture is hidden in the background 
magnetic field $B$. By introducing a very small
$B$ we can control the behavior like we did in the original case 
using $\theta={\rm Im\ }\tau/{\rm Re\ }\tau$ 
(under T-duality $\theta \leftrightarrow {V \over l_s^2} B$).

\begin{figure}
\begin{center}
\leavevmode
\epsfbox{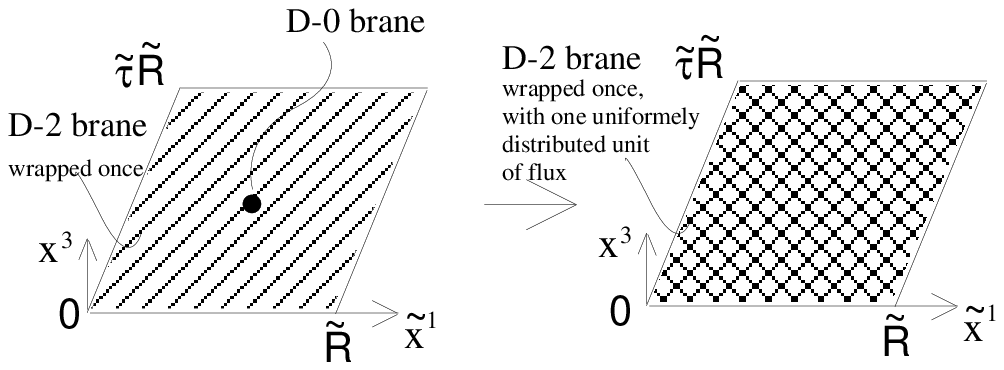}
\end{center}
\caption{Evolution of the initial configuration on the dual torus $\tilde T^2$.
\label{torus2}}
\end{figure}

\newsec{Final Remarks.}

\exclude{
In this conclusion we return to the original problem of D-branes at angles 
but not touching each other.
The method used in this paper to find the conditions for tachyonic 
excitation can be applied in a straightforward manner to the
higher dimensional objects, Dirichlet $p-$branes. And one can 
immediately see the crucial difference between $(1+1)$-dimensional 
and $(p+1)$-dimensional worldvolumes. We would like to claim that 
in the case of 
$p>1$ tachyon always appears, independently of any relative 
distance parameter $a$.
The reason for that is that in the case of $(1+1)$-dimensional field 
theory, the massless U(1) gauge field living in the D-string does not 
have any physical modes, and the masses for the physical modes start
at $m^2={a^2/l_s^4}$ -- the closest distance between D-strings. The 
interaction terms
lower this mass, and at some value of $\theta$ it becomes $m^2\leq0$, and 
the tachyon appears. 
For the case of $(p+1)$-dimensional worldsheet, the massless U(1) gauge 
field does have
physical modes, and when it is mixed with massive 'stretched'
modes by an arbitrarily 
small Hermitean bilinear mixing, as we would expect in the generic case
when the supersymmetry is broken, the re-diagonalized mass matrix will 
have negative mass eigenvalue. (The lighter mass goes down and the 
heavier one goes up, as it happens when we turn the Hermitean bilinear 
mixing on).
}

The virtue of the model considered was that we were able to study
the decay process involving tachyonic excitation explicitly, and
one can perform numerical computations in principle, to find an 
exact shape of the final state. We understand that our example of 
the {\bf Gauge-B} was chosen arbitrarily, and perhaps it is not the best 
one. But using it we can predict an asymptotic behavior of the final 
state.

We did not consider here any contribution from fermionic fields. 
We have assumed superstring 
theory as a starting point for our calculations, so that we did not have
a bosonic tachyon apriori (with a mass of order $1/l_s$). 

\noindent
{\bf Acknowledgments.}

This problem appeared initially as a result of the private conversation 
between M.R.Douglas, J.Maldacena and S.Shenker.
We would like to thank M.R.Douglas for his guidance through this problem,
G.T.Gabadadze for many useful discussions, S.Lukyanov, E.Warnicke 
and L.Motl for some help.


\vskip .3in
\begin{thebibliography}{6666}
\bibitem{D orig}J.Dai, R.G.Leigh and J.Polchinski, Mod. Phys. Lett {\bf A4} (1989) 2073.
\bibitem{witten}E.Witten,  ``Bound States of Strings and $p$-Branes'' hep-th/9510135. 
\bibitem{polch}J.Polchinski, S.Chaudhuri, C.V.Johnson, ``Notes on D-Branes`` hep-th/9602052.
\bibitem{polch2}J.Polchinski, ``TASI Lectures on D-Branes'' hep-th/9611050
\bibitem{bachas}C.Bachas, ``(Half-) a Lecture on D-branes'' hep-th/9701019
\bibitem{taylor}W.Taylor, ``Lectures on D-branes, Gauge Theory and M(atrices)'' hep-th/9801182
\bibitem{CM}C. Callan and J. Maldacena, ``Brane Dynamics from the Born-Infeld action'' hep-th/9708147.
\bibitem{LPT}S.Lee, A.Peet and L.Thorlacius, ``Brane-Waves and Strings'' hep-th/9710097.
\bibitem{angles}M.Berkooz, M.Douglas, R.Leigh, ``Branes Intersecting at Angles,'' hep-th/9606139.
\bibitem{townsend}P.Townsend, ``M-branes at angles'' hep-th/9708074
\bibitem{alec}A.Matusis, ``Interaction of non-parallel D1-branes'' hep-th/9707135


\end{thebibliography}
\end{document}